\documentclass[fleqn,twoside]{article}
\usepackage{espcrc2}
\usepackage{graphicx}
\usepackage[figuresright]{rotating}

\newcommand{\AmS}{{\protect\the\textfont2
  A\kern-.1667em\lower.5ex\hbox{M}\kern-.125emS}}
\title{Six years of ScoX-1 monitoring with BeppoSAX Wide Field Cameras}

\author{P. Santolamazza\address{ASI Science Data Center c/o ESA/ESRIN, via G. Galilei, Frascati,  Italy}, 
        F. Fiore\address[OAR]{INAF- Osservatorio Astronomico di Roma, via di Frascati 33, I-00040 Monteporzio, Italy}, 
        L. Burderi\addressmark[OAR],
	T. Di Salvo\address{Astronomical Institute ``Anton Pannekoek'', University of Amsterdam, The Netherlands} }
\begin{document}

\begin{abstract}
We performed a systematic analysis of 54 Wide Field Camera (WFC) observations of ScoX-1 available in the BeppoSAX public archive. Observations span over the six years of BeppoSAX mission lifetime and include 690 hr of data. 
We searched for shifts and shape changes of the Z pattern in the color-color diagrams. We find that the Z pattern occupies most of the time the same locus in the color-color diagram. There are however a few exceptions, which are discussed in detail.

\vspace{1pc}
\end{abstract}
\maketitle

\section{Introduction}

Sco X-1 is a low mass X-ray binary (LMXB), is the brightest extrasolar X-ray source in the sky and was discovered by R. Giacconi and collaborators in the seminal X-ray rocket experiment of June 1962.
Neutron star LMXB are divided into two classes based upon the different morphology of their tracks in color-color diagrams (CD). The three branches of Z tracks are named  Horizontal (HB), Normal (NB) and Flaring (FB), while  atoll sources are characterized by an ``island'' and a ``banana'' state \cite{HvdK89}.
Track morphology is due to spectral variations on timescales of weeks, days or hours. In the case of ScoX-1, the complete Z track is generally traced out in a few hours to a day. Such variations are interpreted as an accretion sequence \cite{H89}, where the mass accretion rate $\dot{M}$ is assumed to change along the track (see however \cite{bcbc03}). 
Secular shifts and shape changes of the Z track  were reported for several Z sources such as CygX-2 \cite{Kuu96}, whose shifts were interpreted in terms of occultation of the emitting region by a precessing accretion disk,  or very recently for LMC X-2 \cite{shk03}, in which secular shifts of 2.5-10\% are seen. Conversely significant secular variations in the Z track of ScoX-1 were never observed \cite{dvdk00}. However an inclination of 45 deg and a variable disk thickness could produce the same effect, besides reducing the HB extent, as discussed in \cite{Foma01}.

\section{Observations and Data Reduction}

We present 54 BeppoSAX/WFC \cite{j97} observations of the Z source ScoX-1 extracted from the BeppoSAX public archive. WFC are coded mask \cite{jean} cameras with 40$^{o}$x40$^{o}$ total field of view (FOV). Data reduction was performed with the WFC software release 204.204.02. To analyze ScoX-1 spectral variability we define five energy bands: 2.0-3.5 keV; 3.5-6.4 keV; 6.4-9.5 keV; 9.5-16.4 keV; 1.7-19.1 keV, and  two colors. The soft color is the ratio of the 3.5-6.4 keV band to the 2.0-3.5 keV band and hard color is the ratios of the 9.5-16.4 keV band to the 6.4-9.5 keV band.
Colors are built from light curves with a 300 s time binning.

The response of each WFC changed during the six year satellite lifetime and is  also a function of the source position in the FOV. 
In order to study colors and intensity variations we must scale the observed count rates to common position and epoch, that we chose to be the central quadrant of WFC1 at January 2002. The scaling was performed by using ratios of the expected count rates in the above defined energy bands. Count rates were obtained by convolving a model spectrum with the response matrix corresponding to the relevant position in the FOV and epoch. The model spectrum was chosen as close as possible to that of the source to be analyzed. 

\begin{figure}[tb]
%\vspace{-1cm}
%\hspace{-.3cm}
\includegraphics*[keepaspectratio, scale= 0.37]{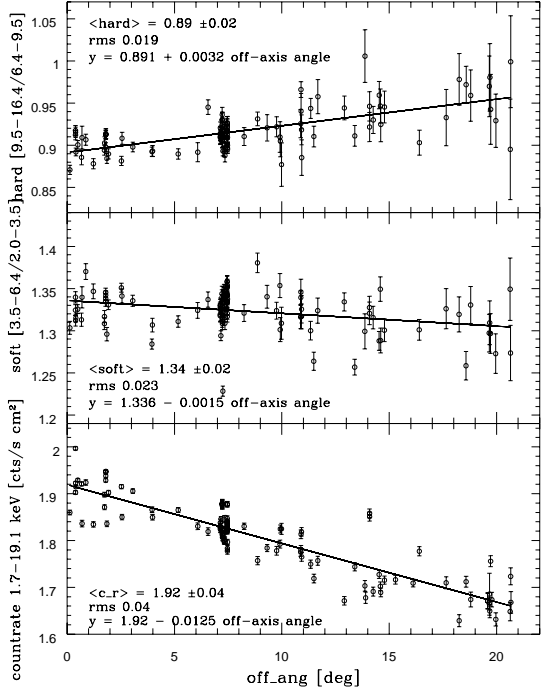}
\vspace{-1cm}
\caption{Crab Nebula colors and intensity vs off-axis angle. From the top: hard color, soft color, count rate in the 2.0-19 keV band }
\vspace{-.6cm}
\end{figure}

To test this procedure we studied colors and intensity of the Crab Nebula, assumed to be a steady X-ray source, in more than 100 observations in which the source was observed at different off-axis angles. 
While we did not find any trend with time we found a residual trend with the off-axis angle in both colors and intensity (Fig.1). An empirical correction for this trend is extimated through a least square fit. The deviations at off-axis angles of 10 and 20 deg are 3.6\% and 7\% for the hard color, 1.1\% and 2.2\% for the soft color, 6.5\% and 13\%  for the intensity. Since the soft color deviation is always smaller  than the statistical error we decided not to correct this color. The corrected CD for all Crab Nebula observations in both cameras are reported in Fig.2, the data from the two camera overlap. The spread in both colors is consistent with a gaussian distribution due to data errors and the  mean hard color and soft color and their standard deviations are 0.89(0.02) and 1.32(0.02), respectively.
The same off-axis angle corrections were applied to ScoX-1 hard color and intensity. 
%Any residual effects  due to a spectral dependance of the  off-axis trend was neglected. 
ScoX-1 count rates were scaled to a common position and epoch by assuming a black body plus a Comptonization model (Xspec {\bf compTT} model)  with spherical geometry.
Model parameter values were fixed to represent a tipical ScoX-1 spectrum at the transition point from NB to FB (hereafter vertex): BB [ kT = 1.35 keV, fl$_{2-10}$ = 0.97$\cdot10^{-7}$ ergs/cm$^{-2}$s$^{-1}$]; COMPTT [ TO = 0.36 keV, kT = 3.2 keV, $\tau$ = 10, fl$_{2-10}$ = 1.38$\cdot10^{-7}$ ergs/cm$^{-2}$s$^{-1}$]. The Galactic column density was set to the value 3$\cdot$10$^{21}$ cm$^{-2}$ as in \cite{bcbc03}. We will give a more detailed description of corrections in \cite{sfbds03}.

\begin{figure}[h]
\vspace{-.7cm}
\hspace{.3cm}
\includegraphics*[ scale= 0.8]{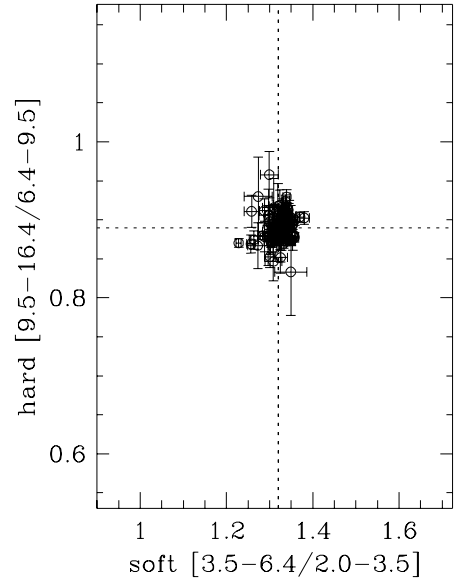}
\vspace{-.9cm}
\caption{Crab Nebula CD on the same scale as following CD of ScoX-1. Colors are scaled to the center of WFC1, epoch January 2002, and corrected for the off-axis angle.}
\vspace{-1.1cm}
\end{figure}

\section{Results}

The light curve of the 1.7-19.1 keV count rate in each WFC observation is shown in Fig.3. Points of count rate $>$ 27.5 come from observations covering the FB only, which is far more luminous than other branches. Conversely other points represent averages over different branches. Observations last from 10 ks to 160 ks and the net exposure is usually 30\%-40\% of the duration. 
To compare observations we divided the data in 9 CD according to the date [1996-97, 1998-99, 2000-01] and to the off-axis angle [5-14 deg, 14-17deg and  17-20 deg]. Most observations have vertices at colors about soft=1.5, hard=0.47 and FB passing through soft=1.8, hard=0.54, see e.g. observations 00001061 in Fig.4a, 20947008 in Fig.4b, 20143001 in Fig.4d, and all observations in Fig.4h. This we call the more common state. However we note some exceptions: observation 20274001 in Fig.4a reaches very low values in both soft color and hard color, its vertex and FB colors are  about 6-8\% lower than the common state location of NB/FB branches; in Fig.4b the vertex and NB of observation 60539002 locate at colors lower by about 4\%  compared with the common state location. 
These shifts indicate secular variations of the track location on timescales of months to years. 
\begin{figure}[tb]
\vspace{-.3cm}
\hspace{-.5cm}
\includegraphics*[keepaspectratio, angle=270, scale=.3]{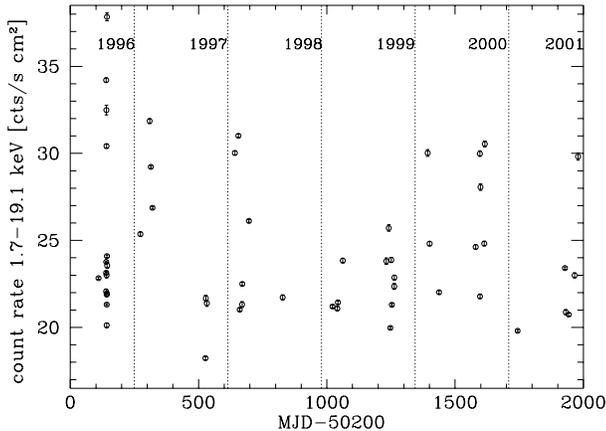}
\vspace{-1cm}
\caption{ Light curve of the total count rate per observation in the 1.7-19.1 keV band }
\vspace{-.8cm}
\end{figure}
Moreover in Fig.4d the lowest points of observations 00001054 and 00001076 lie below the common state vertex location by about 5\% in soft color and 13\% in hard color. These observations are part of the only available nearly continuous monitoring of ScoX-1 that lasts more than 160 ks, from September 13 to 17 in 1996. These observations are plotted on different panels because of their different off-axis angles; we checked that the scaling procedure and off-axis angle correction work well by reporting monitoring observations at different off-axis angles on a same plot and verifying that they build up at least three complete Z pattern consistently. These Z tracks show a maximum shift of 13\% in soft color and 6\%  in hard color among them, suggesting that the track location can also vary on timescales of days.
In Fig.4g and Fig.4h the point spread is wide because errors are larger at these off-axis angles, but the general finding of a common state holds. In Fig.4i the superposition of all observations is shifted to slightly lower values of hard color.
Finally we notice that observation 613150011 in Fig.4c is peculiar, tracing out the HB and NB but never reaching the FB though the duration is about 160 ksec. Note that in other long observations, such as observation 20274001 in Fig.4a, the source moves from HB to the FB and back at least twice on a comparable time.
  
\vspace{-.2cm}

\section{ Conclusions }

Analysis of ScoX-1 CD, spanning over the six years of BeppoSAX lifetime, show that most of the time the track covers the same position. However there are few noticeable exceptions. The most evident is a shift of the vertex and FB positions towards soft color and hard color 6-13\% lower than the more common position. This shifts are observed on timescales of years or even months. Moreover we find indications that shifts appear also on shorter timescale (days). Detailed analysis is in progress to confirm this latter point.

\newpage
\begin{figure}[tbh]
\vspace{-1cm}
\includegraphics*[keepaspectratio, scale=.8]{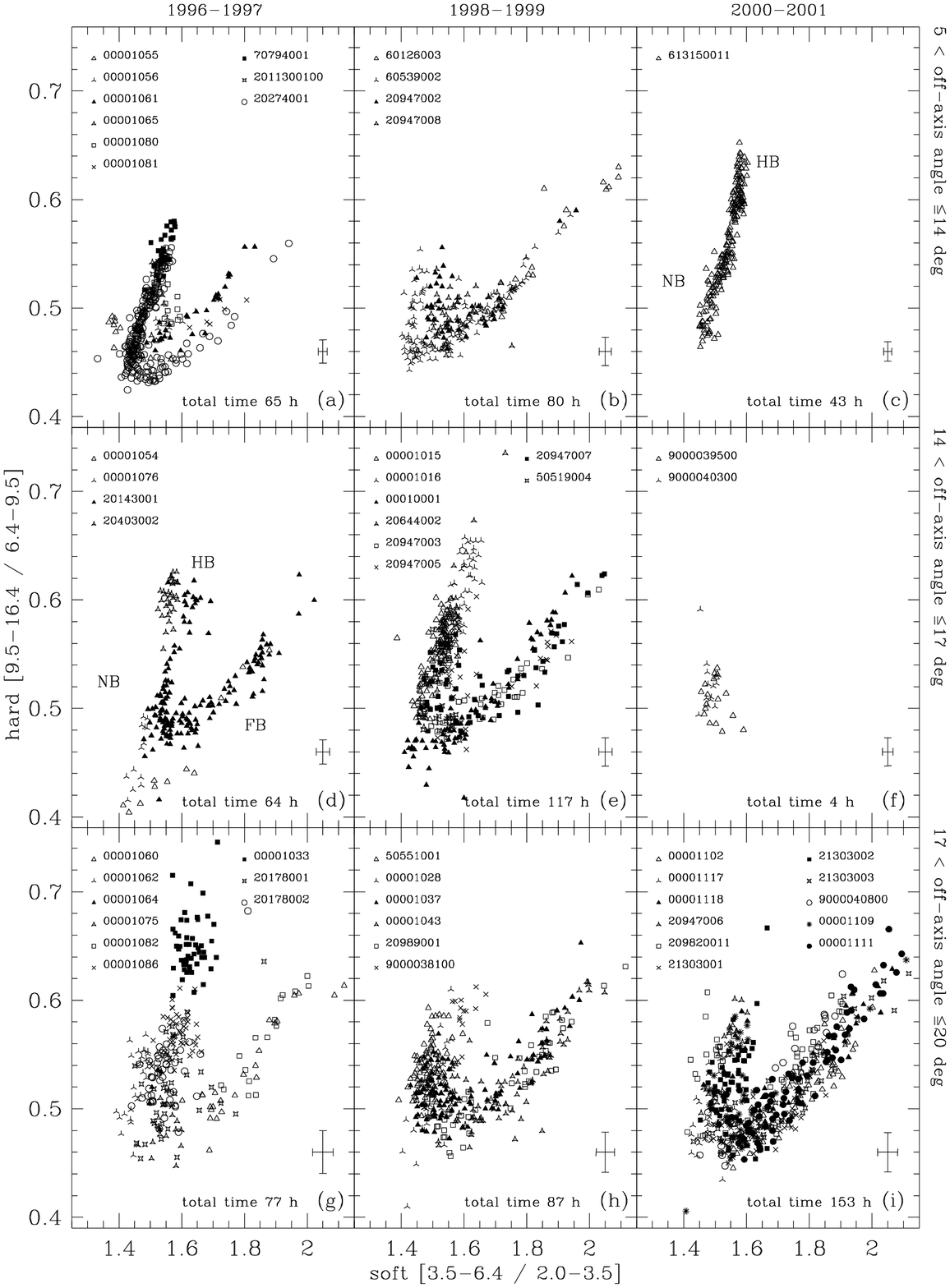}
\vspace{-1.5cm}
\onecolumn
\caption{CD of ScoX-1 grouped according to the date and to the off-axis angle. Observation codes are the same of simultaneous NFI observations in BeppoSAX archive. Tipical error bars and total time for each group of observations are reported.}
\end{figure}
\twocolumn

\end{document}